\documentclass[
notitlepage,
aps,
nobibnotes,
nofootinbib]{revtex4-1}

\usepackage{amsmath}    
\usepackage{mathtools}
\usepackage{upgreek}
\usepackage{bm}         
\usepackage{graphicx}   
\usepackage{grffile}
\usepackage{fancyvrb}   
\usepackage{color}      
\usepackage{subfigure}  
\usepackage{hyperref}   
\allowdisplaybreaks[1]
\usepackage[margin=0.7in]{geometry}
\def\d{\mathrm{d}}
\def\pd{\partial}
\def\b#1{\mathbf{#1}}

\usepackage{wrapfig}
\usepackage{titlesec}
\usepackage{amssymb}

\begin{document}
\author{Matt Majic} \email{mattmajic@gmail.com}
\affiliation{The MacDiarmid Institute for Advanced Materials and Nanotechnology,
School of Chemical and Physical Sciences, Victoria University of Wellington,
PO Box 600, Wellington 6140, New Zealand}
\title{Exact gravitational potential of a homogeneous torus in toroidal coordinates and a surface integral approach to Poisson's equation}

\begin{abstract}
New exact solutions are derived for the gravitational potential inside and outside a homogeneous torus as rapidly converging series of toroidal harmonics. The approach consists of splitting the internal potential into a known solution to Poisson's equation plus some solution to Laplace's equation. The full solutions are then obtained using two equivalent methods, applying differential boundary conditions at the surface, or evaluating a surface integral derived from Green's third identity. This surface integral may not have been published before and is general to all geometries and volume density distributions, reducing the problem for the gravitational potential of any object from a volume to a surface integral.
\end{abstract}

\maketitle
\section{Introduction}
Astronomical toroidal structures have had renewed interest in recent years (see \cite{bannikova2011gravitational} and references therein).
The gravitational potential of a homogeneous torus has been long known as a volume integral of Green's function that appear to have no analytic solution. Series solutions are easier to manipulate, and recently there has been a solution published which expresses the potential in terms of series of solid spherical harmonics with a piecewise expression depending on the region of interest \cite{Kondratyev2009,Kondratyev2010,Kondratyev2012}. Approximate fomulae are given by \cite{bannikova2011gravitational}. However no solution has been found as a series of toroidal harmonics, despite being a natural basis for the problem. Also there does not seem to be a published solution to the exact potential inside the torus (other than a volume integral). 
We find new solutions to the potential inside and outside the torus. In this approach the internal potential is split into a known solution to Poisson's equation plus some solution to Laplace's equation, as done for expample in \cite{hvovzdara2011gravity}. This is analogous to scattering theory where a known external field is incident on a particle. The full solution is found by assuming the potential as a series of toroidal harmonics, and determining the series coefficients by applying boundary conditions of continuity of the potential and its gradient on the surface.

We also derive an integral equation approach to solving the problem, by using Green's third identity in combination with the separation of the internal potential and the boundary conditions to express the solution in terms of a surface integral. This surface integral approach can be applied to any closed object with any volume density distribution. We check that for this problem the two methods - the boundary conditions in differential form and the surface integral produce identical results. The series are shown to converge quickly in all space.

\section{Toroidal coordinates and harmonics}
First define spherical and cylindrical coordinates\footnote{atan2 is a similar to the arctangent but provides correct results in all four quadrants of $x$ and $y$.}:
\begin{align}
r=\sqrt{x^2+y^2+z^2}, \qquad \rho=\sqrt{x^2+y^2}, \qquad \theta=\text{acos}\frac{z}{r}, \qquad \phi=\text{atan2}(y,x) 
\end{align}
Then toroidal coordinates $(\eta,\sigma,\phi)$ with focal ring radius $a$ are defined as 
\begin{align}
\eta=\frac{1}{2}\log\frac{(\rho+a)^2+z^2}{(\rho-a)^2+z^2}, \qquad 
\sigma=\text{sign}(z)\text{acos}\frac{r^2-a^2}{\sqrt{(r^2+a^2)^2-4\rho^2a^2}},
\end{align} 
with $\eta\in[0,\infty) ,~\sigma\in[-\pi,\pi]$. $\eta$ corresponds to the torus size and $\sigma$ to the angle around the minor axis. For convenience we also define:
\begin{align}
\beta&=\cosh\eta=\frac{r^2+a^2}{\sqrt{(r^2+a^2)^2-4\rho^2a^2}}
\end{align}
with $\beta\in[1,\infty)$. \\

Laplace's equation is partially separable in toroidal coordinates, meaning that solutions can be written as a product of functions of each coordinate but must also be multiplied by a coordinate-dependent prefactor. In general, toroidal harmonics are
\begin{align}
\Delta \left\{ { P_{n-1/2}^m(\beta) \atop Q_{n-1/2}^m(\beta) }\right\}   \left\{ {\cos(n\sigma) \atop\sin n\sigma }\right\}\left\{e^{+ im\phi}\atop e^{- im\phi}\right\} , \qquad \Delta=\sqrt{2(\beta-\cos\sigma)} 
\end{align}
where the curly braces indicate any linear combination of their interior functions. However, due to the symmetry of the problem, only functions with $\cos(n\sigma)$ and $m=0$ apply. $P_{n-1/2}^m$ and $Q_{n-1/2}^m$ are Legendre functions of half-integer degree, also called toroidal functions. $P_{n-1/2}^m(\beta)$ are singular on the focal ring, and $Q_{n-1/2}^m(\beta)$ are singular on the entire $z$ axis. Matlab codes to evaluate these are attached as supplementary material.

\begin{figure}
\includegraphics[scale=.4,trim={3cm 4cm 2cm 4cm},clip]{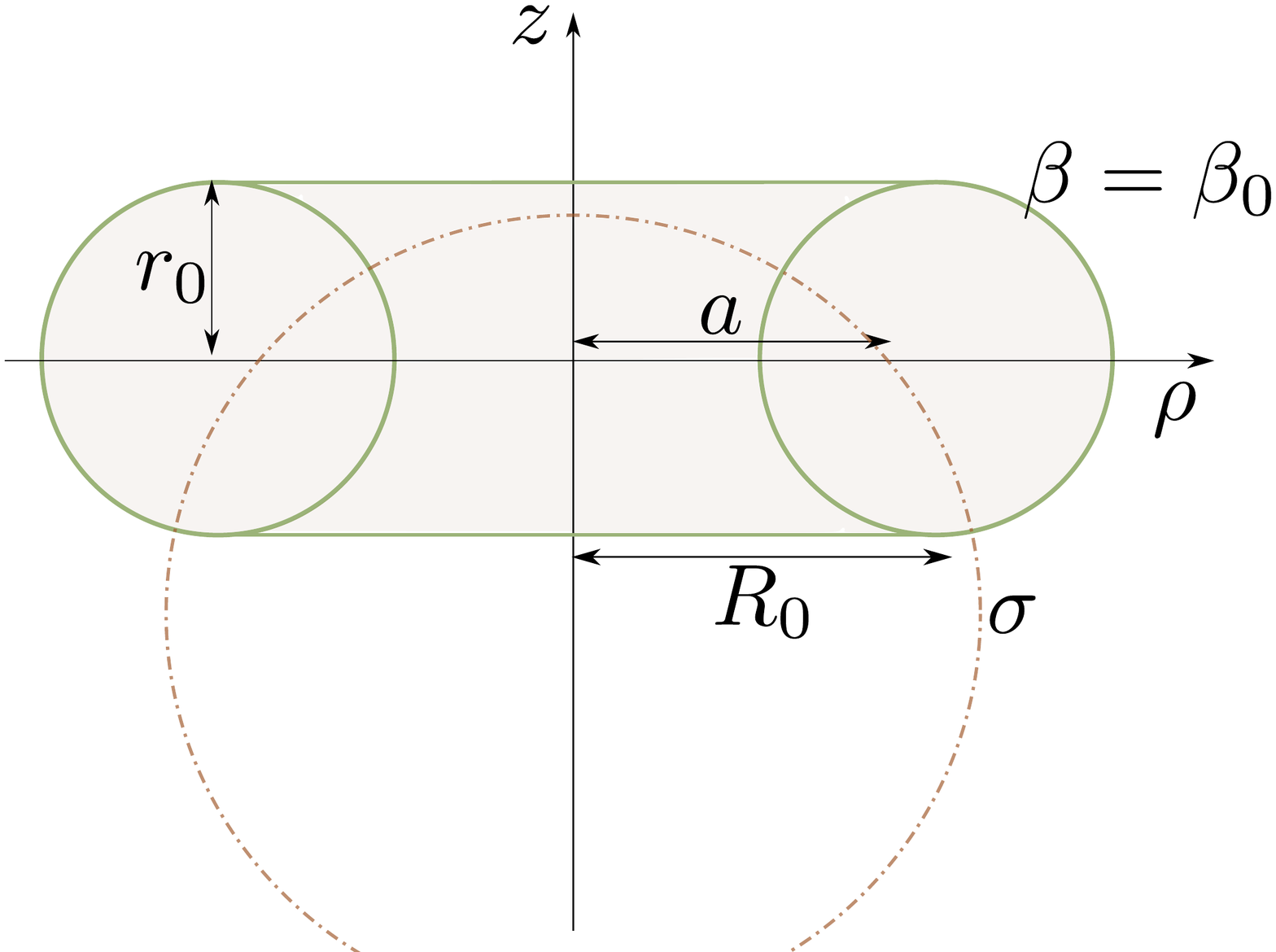}
\caption{Schematic of the coordinates used. $\beta$ defines the torus size and $\sigma$ relates to the angle around the minor axis.} \label{torus schem}
\end{figure}

\section{Problem}
\subsection{Derivation of solution}
Consider a torus of uniform density $\tau$, with major radius $R_0$ and minor radius $r_0$ as shown in figure \ref{torus schem}. The focal ring  radius $a$ and the surface parameter $\beta=\beta_0$ can be obtained through
\begin{align}
a=\sqrt{R_0^2-r_0^2} \qquad \beta_0=\frac{R_0}{r_0}
\end{align}   
Denote the potential inside the torus as $V_i$ and outside as $V_o$.
Inside the torus we split the potential into two parts $V_i=V_P+V_L$ where $V_P$ satisfies Poisson's equation and $V_L$ satisfies Laplace's equation. So we have 
\begin{align}
\nabla^2 V_P&=-4\pi G \tau \label{Poisson}\\
\nabla^2 V_L&= \nabla^2 V_o=0.
\end{align}
The boundary conditions are continuity of the potential and continuity of the gradient of the potential, equivalent to the gravitational acceleration field being finite and continuous. In particular at the surface:
\begin{align}
V_i=V_o \qquad \partial_\beta V_i=\partial_\beta V_o \qquad \text{at } \beta=\beta_0 \label{BCs}
\end{align} 
We formulate the problem similar to that of a scattering problem in electromagnetism or acoustics: assuming $V_P$ is known, we want to find $V_L$ and $V_o$ that satisfy the boundary conditions. It is easy to find solutions to \eqref{Poisson}, for example $V_P\propto x^2$ or $V_P\propto r^2$ are both solutions. The choice of $V_P$ is not unique; any multiple of $V_L$ can be added to $V_P$ and it will still solve Poisson's equation.

For this problem we found it convenient to choose
\begin{align}
V_P=-\frac{2}{3}\pi G \tau (r^2+a^2).
\end{align}
Now we assume the potentials can be expressed as series of toroidal harmonics. This cannot be done exactly for $V_P$, but it can still be expressed in a similar fashion. We have
\begin{align}
V_P&=\Delta\sum_{n=0} a_n(\beta) \cos(n\sigma) \label{VP}\\
V_L&=\Delta\sum_{n=0} b_n Q_{n-1/2}(\beta) \cos(n\sigma) \label{VL}\\
V_o&=\Delta\sum_{n=0} c_n P_{n-1/2}(\beta) \cos(n\sigma) \label{Vo}
\end{align}
To find the coefficients $a_n$, we first express $r^2+a^2$ in toroidal coodinates:
\begin{align}
r^2+a^2=\frac{2a^2\beta}{\beta-\cos\sigma} = \frac{\sqrt{2}a^2\Delta\beta}{(\beta-\cos\sigma)^{3/2}}
\end{align}
then use the identity \cite{Scharstein2005}: 
\begin{align}
\frac{1}{(\beta-\cos\sigma)^{3/2}} = \frac{-2^{3/2}}{\pi}\sum_{n=0}^\infty \epsilon_n Q_{n-1/2}'(\beta)\cos(n\sigma), \qquad \epsilon_n = 
\begin{cases}
1 & n=0 \\ 2 & n\neq 0 \label{an}
\end{cases}
\end{align}
where the prime denotes differentiation with respect to the argument. So we have
\begin{align}
a_n(\beta)= \frac{8}{3} G \tau a^2 \epsilon_n	\beta Q_{n-1/2}'(\beta)
\end{align}
Now we apply the boundary conditions, and equate coefficients of $\cos(n\sigma)$ to obtain simultaneous equations to solve for $b_n$ and $c_n$. Because $\Delta$ depends on $\beta$, the condition on the derivative appears to give cumbersome equations with many terms. However, there are cancellations and the solutions are quite simple:
\begin{align}
b_n &= \sinh^2\eta_0 [ P_{n-1/2}(\beta_0) a_n'(\beta_0) - P_{n-1/2}'(\beta_0) a_n(\beta_0) ] \label{bn}\\
c_n &= \sinh^2\eta_0 [ Q_{n-1/2}(\beta_0) a_n'(\beta_0) - Q_{n-1/2}'(\beta_0) a_n(\beta_0) ] \label{cn}
\end{align}
Despite the fact that the choice of $a_n$ is not unique and may differ by any multiple of $d_nQ_{n-1/2}(\beta)$ for constant $d_n$, it is straightforward to show this arbitrary choice has no effect on $V_i$ or $V_o$.

\subsection{Integral equation approach}
The problem can also be expressed in terms of surface integrals. First, Green's third identity applied to the inside volume for $V_L$ and the outside volume for $V_o$ are:
\begin{align}
\int_S   V_L(\mathbf{r'}) \frac{\partial G(\mathbf{r},\mathbf{r'})}{\partial n'} - G(\mathbf{r},\mathbf{r'}) \frac{\partial V_L(\mathbf{r'})}{\partial n'} \mathrm{d}s' = 
\begin{cases} 
-V_L(\mathbf{r}) &\mathbf{r}\in \text{torus}\\ 
0 &\mathbf{r}\notin \text{torus} \end{cases} \\
\int_S V_o(\mathbf{r'}) \frac{\partial G(\mathbf{r},\mathbf{r'})}{\partial n'} - G(\mathbf{r},\mathbf{r'}) \frac{\partial V_o(\mathbf{r'})}{\partial n'} \mathrm{d}s' = 
\begin{cases} 
0 \quad &~~~\mathbf{r}\in \text{torus}\\ 
V_o(\mathbf{r}) &~~~ \mathbf{r}\notin \text{torus}\end{cases}
\end{align}
where $G(\mathbf{r},\mathbf{r'})$ is Green's function, $\mathrm{d}s'$ is a surface element and $\b n'$ is the unit normal to the surface, pointing outwards. Subtracting these two equations and using the boundary conditions \eqref{BCs} we find
\begin{align}
\int_S V_P(\mathbf{r'}) \frac{\partial G(\mathbf{r},\mathbf{r'})}{\partial n'} - G(\mathbf{r},\mathbf{r'}) \frac{\partial V_P(\mathbf{r'})}{\partial n'} \mathrm{d}s' = 
\begin{cases} 
V_L \quad &\mathbf{r}\in \text{torus}\\ 
V_o(\mathbf{r}) &\mathbf{r}\notin \text{torus}\end{cases} \label{int eqs}
\end{align} 
which gives the potential explicitly if $V_P$ is known.
In order to use these equations we again expand everything as series, including Green's function \cite{Morse1953}:
\begin{align}
G(\mathbf{r},\mathbf{r'})=\frac{1}{4\pi|\b r-\b r'|} 
&=\frac{\Delta\Delta'}{8\pi^2 a}\sum_{m=0}^\infty \sum_{n=0}^\infty\epsilon_n\epsilon_m  P_{n-1/2}^m(\beta)Q_{n-1/2}^{-m}(\beta')[\cos(n\sigma)\cos(n\sigma')+\sin(n\sigma)\sin(n\sigma')]\cos m(\phi-\phi') \label{GF} 
\end{align}
Now we insert this along with the series expressions (\ref{VP}-\ref{Vo}) (the series for $V_P$ should use a different summation index to $n$) into the integral equations \eqref{int eqs}, apply the differentiation, and equating coefficients of $\cos(n\sigma)$ on both sides. Here only terms with $m=0$ and $\cos(n\sigma)$ survive the integration. The surface normal and derivative are
\begin{align}
\frac{\pd}{\pd n}=\frac{-\Delta^2\sinh\eta}{2a}\frac{\pd}{\pd\beta}, \qquad 
\d s= \frac{4a^2\sinh\eta}{\Delta^4}\d\sigma\d\phi.
\end{align}
After some simple calculus and algebra we have for $c_n$ (the integration primes have been omitted):
\begin{align}
c_n=\frac{\sinh^2\eta_0}{2\pi}\epsilon_n \sum_{k=0}^\infty \int_{-\pi}^\pi \cos(n\sigma)\cos(k\sigma) \left[ Q_{n-1/2}(\beta_0) \left(a_k'(\beta_0) + \frac{a_k(\beta_0)}{\Delta^2} \right) - a_k(\beta_0)\left( Q_{n-1/2}'(\beta_0) + \frac{Q_{n-1/2}(\beta_0)}{\Delta^2} \right) \right] \d\sigma
\end{align}
Initially this looks complicated but the terms with $1/\Delta^2$ cancel, leaving a simple integration over $\cos(n\sigma)\cos(k\sigma)$ which is zero for $n\neq k$, so this expression reduces to\eqref{cn}. Similarly for $b_n$.

\subsection{Numerical implementation}

We now present computationally friendly forms of the series coefficients. 
The derivatives of the Legendre functions can be computed as 
\begin{align}
Q_{n-1/2}'(\beta)=\frac{Q_{n-1/2}^1(\beta)}{\sinh\eta} \label{dQ}
\end{align}
\eqref{dQ} also applies to $P_{n-1/2}$. \\
$a_n'$ contains double derivatives which can be evaluated using the Legendre differential equation,
so that 
\begin{align}
a_n'(\beta)= \frac{8}{3} G \tau a^2 \epsilon_n \left[ \frac{\beta^2+1}{(\beta^2-1)^{3/2}}Q_{n-1/2}^1(\beta) - \frac{(n^2-1/4)\beta}{\beta^2-1}Q_{n-1/2}(\beta) \right]
\end{align}

\begin{figure}[h]
\includegraphics[scale=.66]{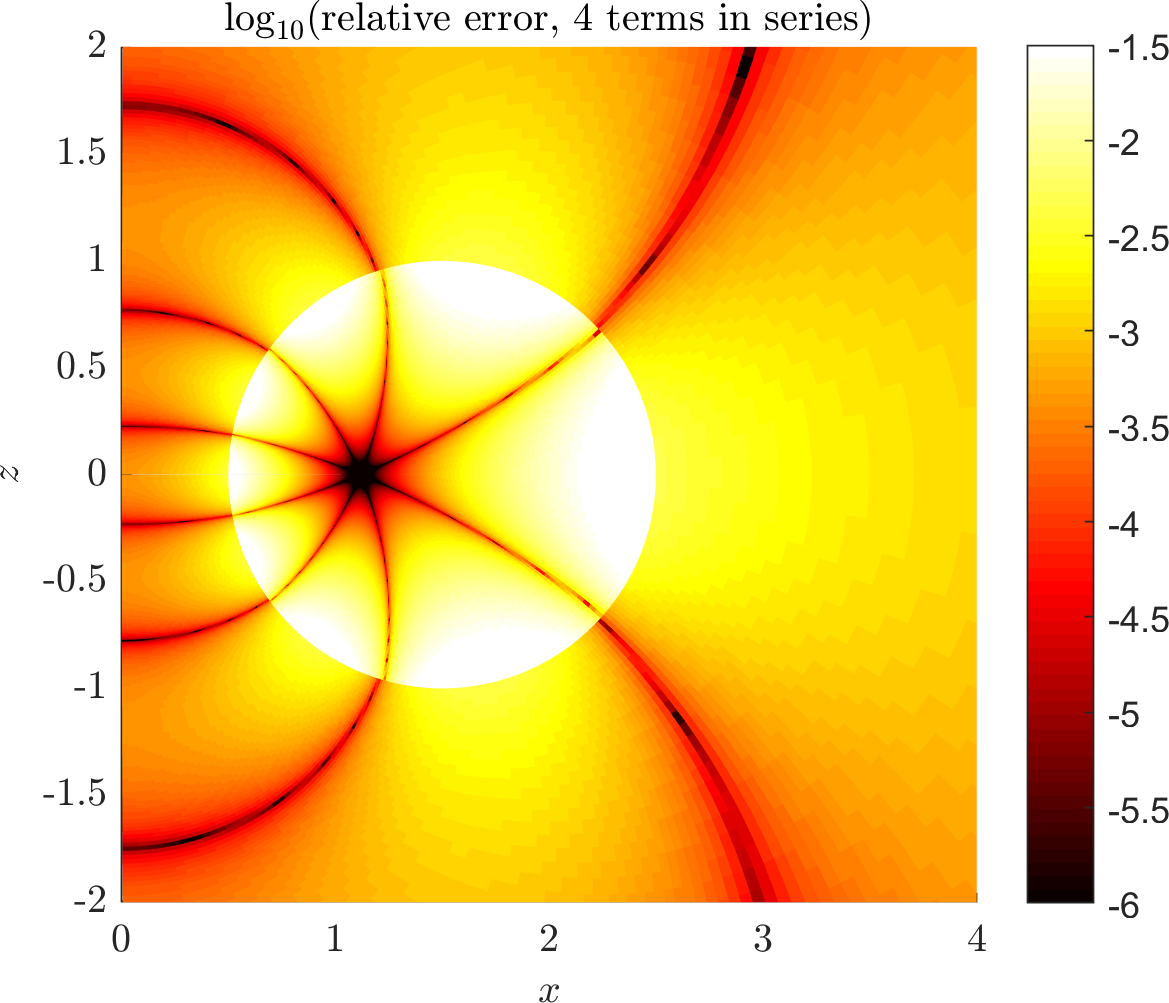}
\includegraphics[scale=.66]{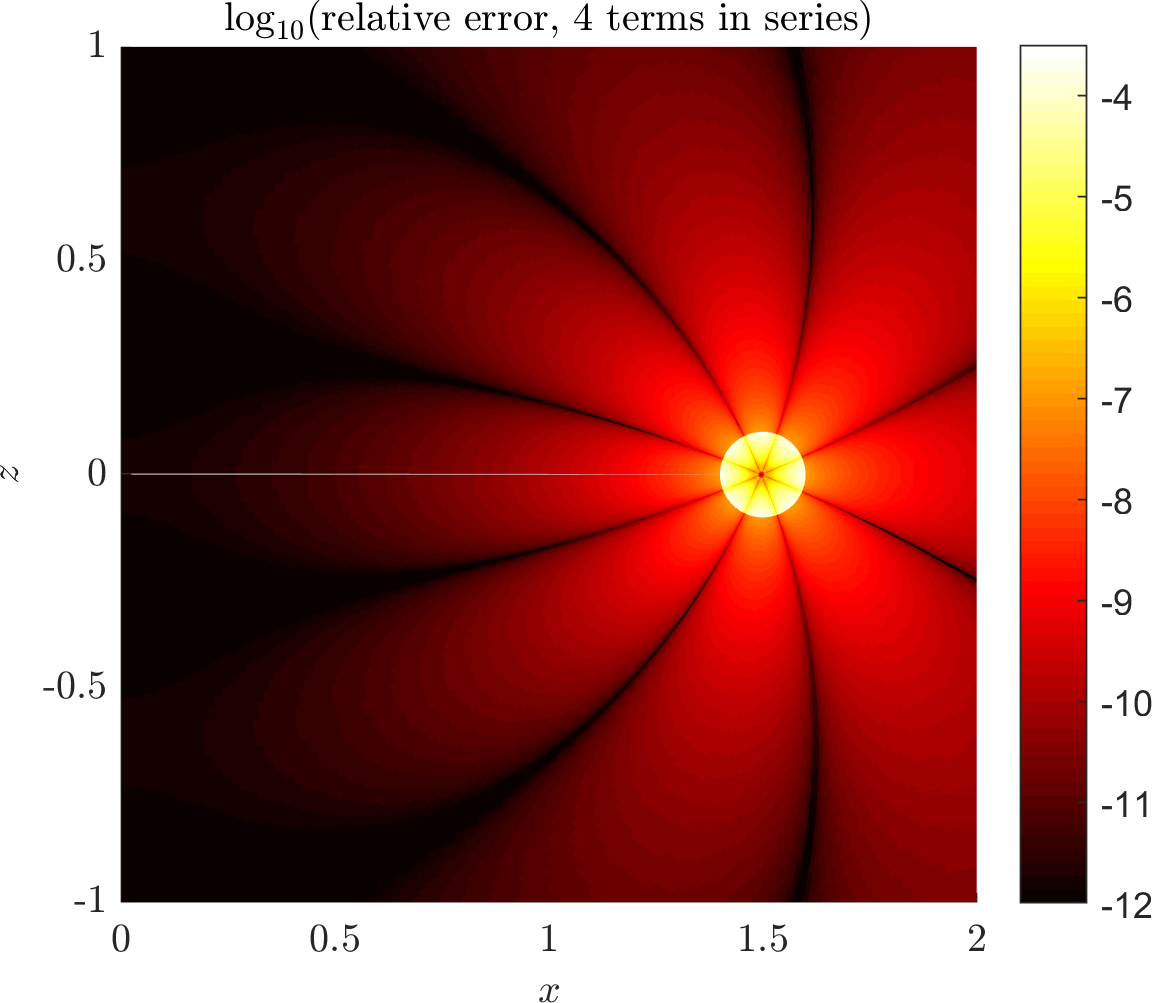}
\caption{Relative error in computing the potential with 4 terms in the series. On the left, $R_0=1.5$, $r_0=1$ and on the right $R_0=1.5$, $r_0=0.1$. The series is compared to itself computed with a higher number of terms. The thin dark lines are where the absolute error passes through zero.}
\label{relerr}
\end{figure}

The toroidal series for $V_o$ has been checked against the spherical series in the outer region ($r>R_0$) \cite{Kondratyev2009}. As shown in figure \ref{relerr}, the toroidal series can be computed accurately with only summing up to $n=3$. The slower convergence inside the torus may be due to an inefficient choice of $V_P$, which means this aspect could be improved.

The near and far field spherical harmonic series' also converge rapidly except near the  spherical annulus $a<r<R_0$. The series for $r>R_0$ \cite{Kondratyev2009} has the advantage over the toroidal series in that it doesn't contain any special functions so is much easier to evaluate, but the series for $r<r_0$ \cite{Kondratyev2010} and $a<r<R_0$ \cite{Kondratyev2012} contain elliptic integrals or hypergeometric functions.

\subsection{Acceleration vector}
For convenience we present the acceleration vector $\b g$. First, in terms of toroidal unit vectors:
\begin{align}
\b g =  \nabla V &= \frac{\beta-\cos\sigma}{a}\left[\frac{\pd V}{\pd\eta}\hat{\bm\upeta} + \frac{\pd V}{\pd\sigma}\hat{\bm\upsigma} \right] 
\end{align}
For example consider the potential outside the torus. The derivatives are
\begin{align}
\frac{\pd V_o}{\pd \eta} &= \frac{\sinh\eta}{\Delta^2}V_o + \Delta\sum_{n=0}^\infty c_n P_{n-1/2}^1(\beta) \cos(n\sigma) \\
\frac{\pd V_o}{\pd \sigma} &= \frac{\sin\sigma}{\Delta^2}V_o + \Delta\sum_{n=0}^\infty c_n P_{n-1/2}(\beta) n\sin(n\sigma)
\end{align}

And for cylindrical unit vectors:
\begin{align}
\b g &= \frac{\pd V}{\pd\rho} \hat{\bm\uprho} + \frac{\pd V}{\pd z} \hat{\b z} \\
&=\left(\frac{\pd\eta}{\pd\rho}\frac{\pd V}{\pd\eta} + \frac{\pd\sigma}{\pd\rho}\frac{\pd V}{\pd\sigma}\right) \hat{\bm\uprho} + \left(\frac{\pd\eta}{\pd z}\frac{\pd V}{\pd\eta} + \frac{\pd\sigma}{\pd z}\frac{\pd V}{\pd\sigma} \right)\hat{\b z}
\end{align}
The partial derivatives are \cite{Fukushima2016}
\begin{align}
\frac{\pd\eta}{\pd\rho}&=\frac{a^2-\rho^2+z^2}{2a^4}\rho (\beta-\cos\sigma)^3\sinh\eta \\
\frac{\pd\eta}{\pd z}&=\frac{-\rho^2z}{a^4}(\beta-\cos\sigma)^3\sinh\eta \\
\frac{\pd\sigma}{\pd\rho}&=\frac{-\rho z}{a^3} (\beta-\cos\sigma)^2 \\
\frac{\pd\sigma}{\pd z}&=\frac{\rho^2-z^2-a^2}{2a^3} (\beta-\cos\sigma)^2
\end{align}

\section{Conclusion}
We have derived a toroidal harmonic solution to the gravitational field of a homogeneous torus, using either differential boundary conditions or surface integral equations. These surface integrals could replace the standard volume integral equations for an object of any shape or density distribution, so long as some particular solution to Poisson's equation can be found. This atleast reduces the number of integrals required to evaluate by one. 

In reality, astronomical toroidal structures would not be uniformly distributed, and it would be ideal to find solutions for simple density distributions, possibly $\beta$ dependent. Other shapes could be treated with this method, for example tori with elliptical cross sections, parameterised with flat-ring cyclide coordinates \cite{Moon1988}. \\

\bigskip
\textbf{Acknowledgments.} This research was funded by a Victoria University of Wellington doctoral scholarship. Thanks to Eric C. Le Ru for helpful suggestions.

\bibliography{../library, ../Tmatrix, ../references}

\end{document}